\documentclass[a4paper, 11pt]{article}          % Paper and font size
\usepackage[english]{babel}                     % Language setting

\usepackage[top=2cm,bottom=2cm,left=2cm,right=2cm]{geometry}

\usepackage{graphicx}                           % Graphics
\usepackage{amsmath}                            % Math
\usepackage[table]{xcolor}                             % Link color
\usepackage{hyperref}
\hypersetup{colorlinks=true, allcolors=.}
\usepackage[T1]{fontenc}                        % Special characters
\usepackage{fancyhdr}                           % Load header, footer package
\usepackage{textgreek}

\usepackage{lastpage}                           % Footer note
\usepackage{lipsum}                             % For dummy text

\usepackage[singlelinecheck=false,labelfont=bf]{caption}
\usepackage{subcaption}

\usepackage[bottom]{footmisc}

\usepackage{authblk}

\usepackage{multirow}

\usepackage{abstract}
\begin{document}

\sloppy

\setlength{\parskip}{12pt}

% Figure
%\begin{figure}[h]
%  \centering
%    \includegraphics{}
%    \caption{}\label{fig:1}
%\end{figure}

%\begin{table}[h]
%\caption{}\label{tab:1}
%\begin{tabular}{ll}
%\hline
%  &  \\ 
%\midrule
% & \\
% & \\
% & \\
% & \\
%\hline
%\end{tabular}
%\end{table}

% Set title, author, date
\title{Towards AMS measurements of \textsuperscript{91}Nb, \textsuperscript{94}Nb and \textsuperscript{93}Mo\\produced in fusion environment}

\author[a]{Carlos~Vivo-Vilches}

\author[a]{Esad~Hrnjic}

\author[a]{Martin~Martschini}

\author[a]{Kyra~Altindag}

\author[b]{Lee~W.~Packer}

%\author[a]{Peter~Steier}

\author[a]{Robin~Golser}

\author[a]{Karin~Hain}

\affil[a]{University of Vienna, Faculty of Physics, Isotope Physics, Waehringer~Strasse~17,~1090~Vienna,~Austria}

\affil[b]{United Kingdom Atomic Energy Authority, Culham Campus, Abingdon,~OX14~3DB,~United~Kingdom}

%\affil[c]{International Atomic Energy Agency, Wagramer Strasse 5, 1220 Vienna, Austria}

\affil[*]{Corresponding author. \textit{E-mail:} \href{mailto:carlos.vivo@univie.ac.at}{carlos.vivo@univie.ac.at} (C. Vivo-Vilches)}

\date{}

\pagenumbering{gobble}
\pagenumbering{arabic}

\maketitle

%\hrule

doi:\href{https://doi.org/10.1016/j.nimb.2025.165847}{10.1016/j.nimb.2025.165847}

©This manuscript version is made available under the CC-BY 4.0 license\\
\href{https://creativecommons.org/licenses/by/4.0}{https://creativecommons.org/licenses/by/4.0}%\\

%\hrule

%\newpage
\begin{abstract}
% "does not exceed 250 words"

Long-lived radionuclides, such as \textsuperscript{91}Nb, \textsuperscript{94}Nb and \textsuperscript{93}Mo, are expected to be produced in nuclear fusion reactors by reactions of high-energy neutrons with the structural material. Accurate predictions of waste categorization require experimental validation of simulation codes like FISPACT-II. This work explores the use of Ion-Laser InterAction Mass Spectrometry (ILIAMS) at the Vienna Environmental Research Accelerator (VERA) to measure these three radionuclides by accelerator mass spectrometry (AMS).  The ILIAMS setup employs laser photodetachment to suppress their respective stable isobars: \textsuperscript{91}Zr, \textsuperscript{94}Zr and \textsuperscript{94}Mo, and \textsuperscript{93}Nb.

For \textsuperscript{91,94}Nb measurements, NbO\textsubscript{3}\textsuperscript{--} is selected, with interferences from ZrO\textsubscript{3}\textsuperscript{--} suppressed by collisions with the He buffer gas in the ion cooler. The suppression can be enhanced by overlapping a 355\,nm laser with the ion beam. The lower limit for the suppression factor is 37000. In that way, we reach \textsuperscript{91}Zr/\textsuperscript{93}Nb and \textsuperscript{94}Zr/\textsuperscript{93}Nb levels of 1.2\,$\times$\,10\textsuperscript{--14} and 1.8\,$\times$\,10\textsuperscript{--14}, respectively, in targets prepared from commercial Nb\textsubscript{2}O\textsubscript{5}. MoO\textsubscript{3}\textsuperscript{--} is suppressed by a factor of 4360, leading to a \textsuperscript{94}Mo/\textsuperscript{93}Nb interference of 1.28\,$\times$\,10\textsuperscript{--10} in the same targets.

For \textsuperscript{93}Mo measurements, MoO\textsubscript{2}\textsuperscript{--} is selected, with interference from NbO\textsubscript{2}\textsuperscript{--} suppressed by 637\,nm photons by a factor of 5.5\,$\times$\,10\textsuperscript{6}. This results in a \textsuperscript{93}Nb/\textsuperscript{nat}Mo level of 1.3\,$\times$\,10\textsuperscript{--13} in targets prepared from commercial MoO\textsubscript{3}.

Suppression factors as high as this are not achieved by isobar suppression techniques based on differences in energy loss, not even by AMS facilities with terminal voltages above 8.5\,MV.

\end{abstract}

%\newpage

%\hrule

\ 

%\hrule

\section{Introduction}\label{sec:intro}

In future nuclear fusion reactors, radionuclides are expected to be produced by the reactions of the \mbox{high-energy} neutrons (14.1\,MeV) from the deuterium-tritium (D-T) fusion with the nuclei in the structural material of the reactor. Unlike current nuclear fission reactors, fusion reactors are not anticipated to produce long-lived high level waste. However, research is still required to predict for which materials and under which conditions the waste from fusion reactors will be able to be categorized as low level waste or intermediate level waste 100-300~years after the end of operation. These predictions are performed with computer codes for the simulation of activation of materials, such as FISPACT-II \cite{FISPACT}, whose results rely on the accuracy of the reaction cross sections provided by the libraries used, such as TENDL \cite{TENDL-2012,TENDL-2019}.

To validate these codes, experimental campaigns have been conducted at the Joint Experimental Torus (JET) reactor, where foils of different materials intended for use in the ITER reactor were placed inside the plasma chamber. After the campaign the activities of several short lived radionuclides were measured in those foils and compared with the ones calculated with FISPACT-II taking into account the fluence and energy distribution of the neutrons, and the irradiation schedule \cite{Packer2021,Packer2024}. While the weighted average of the ratio between calculated and experimental activities (C/E) for the D-T campaign DTE2 was 0.986\,$\pm$\,0.07, for some radionuclides and/or materials this ratio deviated significantly from 1, highlighting the importance of these experimental campaigns. An example is \textsuperscript{60}Co, where the C/E ratio is 2.09\,$\pm$\,0.04 for stainless steel SS316L and 3.29\,$\pm$\,0.03 for SS316L(N) \cite{Packer2024}. Since these measurements were non-destructive, the neutron irradiated foils, with masses between 0.5 g and 1.0\,g, are still available for the measurement of long-lived radionuclides. In materials containing molybdenum, three long-lived radioisotopes become important contributors to the dose at 1\,m from the source 100-300~years after the end of operation, and therefore of relevance for the long-term disposal of these materials: \textsuperscript{91}Nb, \textsuperscript{94}Nb, and \textsuperscript{93}Mo \cite{Gilbert2020}.

The interaction of D-T neutrons with molybdenum produces \textsuperscript{94}Nb through the \textsuperscript{94}Mo(n,p)\textsuperscript{94}Nb, \textsuperscript{95}Mo(n,np)\textsuperscript{94}Nb and \textsuperscript{95}Mo(n,d)\textsuperscript{94}Nb reactions \cite{Gilbert2020}. This radionuclide decays by \textbeta\textsuperscript{--}-emission to \textsuperscript{94}Mo with a half-life of (20300\,$\pm$\,1600)\,a \cite{Schuman1959}. In 100\% of these decays, two \textgamma\ particles are emitted, one with an energy of 702.65\,keV and another of 871.091\,keV \cite{NuclearData94}. Therefore, it can be detected by conventional \textgamma-spectrometry, which typically does not require the destruction of the sample or extensive sample preparation \cite{Oshima2020}. Detection limits down to 0.05\,Bq can be obtained, though, if the sample is chemically treated to separate \textsuperscript{94}Nb from other \textgamma-emitting radionuclides \cite{Tanaka2014,Shimada2016}.

\textsuperscript{93}Mo, which decays by electron capture to \textsuperscript{93}Nb with a half-life of (4839\,$\pm$\,63)\,a \cite{Kajan2021}, is produced in fusion reactors by the interaction of D-T neutrons with molybdenum through the \textsuperscript{94}Mo(n,2n)\textsuperscript{93}Mo and \textsuperscript{92}Mo(n,\textgamma )\textsuperscript{93}Mo reactions. Its detection is commonly done by liquid scintillation counting (LSC) of Auger electrons \cite{Luo2022} or by detection of the Nb X-Rays emitted after the decay \cite{Shimada2016,Shimada2017}. The latest shows the lowest detection limit found in the literature, 0.02\,Bq~\cite{Shimada2017}. Even using gas collision cells, inductively coupled plasma mass spectrometry (\mbox{ICP-MS}) measurements of \textsuperscript{93}Mo are limited by the molecular interference of \textsuperscript{92}ZrH\textsuperscript{+} and \textsuperscript{92}MoH\textsuperscript{+}, reaching detection limits of 0.2\,Bq in the best case \cite{Do2021}.

The development of the detection of \textsuperscript{94}Nb and \textsuperscript{93}Mo by radiometric techniques up to now was motivated by their production in nuclear fission reactors, due to thermal neutron capture on their respective stable isotopes, \textsuperscript{93}Nb and \textsuperscript{92}Mo. Accordingly, there are hardly any published detection limits for \textsuperscript{91}Nb, which cannot be produced by neutron absorption on any stable nuclide. In contrast, \textsuperscript{91}Nb is produced in fusion reactors by the interaction of D-T neutrons with molybdenum either directly through the \textsuperscript{92}Mo(n,np)\textsuperscript{91}Nb and \textsuperscript{92}Mo(n,d)\textsuperscript{91}Nb  reactions, or indirectly through the \textsuperscript{92}Mo(n,2n)\textsuperscript{91}Mo reaction reaction and subsequent decay of \textsuperscript{91}Mo ($T_{1/2}$\,=\,15.49\,min) \cite{Gilbert2020}. \textsuperscript{91}Nb is expected to be the second most active or even the most active radionuclide in the blanket of nuclear fusion reactors 100~years after the end of operation for stainless steel with more than 1\% of molybdenum \cite{Bailey2021}. The half-life of this radionuclide, which decays to \textsuperscript{91}Zr by either electron capture (99.83\%) or \textbeta\textsuperscript{+} emission (0.17\%) \cite{NuclearData91}, is known only with a high uncertainty, the only published value for it being (680 ± 130)\,a \cite{Nakanashi1982}.

The D-T neutron fluence during the DTE2 campaign at JET was around 10\textsuperscript{15}\,cm\textsuperscript{--2} \cite{Packer2024}. Using the cross sections tabulated in TENDL 2023 \cite{TENDL-2019,TENDL-2023}, the order of magnitude for the expected specific activities of \textsuperscript{91}Nb, \textsuperscript{94}Nb and \textsuperscript{93}Mo in the irradiated material with the highest molybdenum content, Inconel 718 (3.3\%), would be 1\,Bq/g, 0.001\,Bq/g and 0.1\,Bq/g, respectively. Taking into account that the specific activities of \textsuperscript{91}Nb, \textsuperscript{94}Nb and \textsuperscript{93}Mo could also be an order of magnitude lower, the detection limits of radiometric techniques do not allow the measurement of \textsuperscript{91}Nb, \textsuperscript{94}Nb and \textsuperscript{93}Mo in the foils irradiated during the DTE2 campaign at the JET reactor, at least not without destroying a substantial amount of each foil. Since molecular background is a clear limitation for the measurement of these radionuclides by routine mass spectrometry methods, such as ICP MS \cite{Do2021,Russel2021}, accelerator mass spectrometry (AMS) becomes the only possibility for such measurements. Nevertheless, in this mass range it is very challenging to suppress the interferences caused by stable isobars: \textsuperscript{91}Zr in the case of \textsuperscript{91}Nb; \textsuperscript{94}Zr and \textsuperscript{94}Mo in the case of \textsuperscript{94}Nb; and \textsuperscript{93}Nb in the case of \textsuperscript{93}Mo. For these relative differences in the atomic number of \textDelta Z/Z close to 1/100, isobar separation techniques based on the difference in energy loss reach their limit, even at terminal voltages above 8.5\,MV and using foil stripping~\cite{Guozhu2013,Lu2015,Hain2018,Koll2019,Pavetich2022}.

This paper presents the preliminary studies on the capability of the Ion-Laser InterAction Mass Spectrometry (ILIAMS) setup of the 3-MV-AMS facility VERA (Vienna Environmental Research Accelerator) at the University of Vienna \cite{Martschini2017,Martschini2019,Martschini2022} to deal with the stable isobars of \textsuperscript{91}Nb, \textsuperscript{94}Nb, and \textsuperscript{93}Mo. In this setup, isobar suppression is achieved by laser photodetachment of negative ions. ILIAMS has already proven its capabilities for \textsuperscript{36}Cl \cite{Lachner2019}, \textsuperscript{26}Al \cite{Lachner2021}, \textsuperscript{90}Sr \cite{Honda2022}, \textsuperscript{135,137}Cs \cite{Wieser2023} and \textsuperscript{182}Hf \cite{Martschini2020}, while its potential for other radionuclides, such as \textsuperscript{99}Tc, is currently under investigation \cite{Hain2022}. This work focuses on the study of the suppression of \textsuperscript{91,94}ZrO\textsubscript{3}\textsuperscript{--} and \textsuperscript{94}MoO\textsubscript{3}\textsuperscript{--} with a 355\,nm laser for \textsuperscript{91,94}Nb AMS, and the suppression of \textsuperscript{93}NbO\textsubscript{2}\textsuperscript{--} with a 637\,nm laser for \textsuperscript{93}Mo AMS.

\section{Methods}\label{sec:methods}

\subsection{The ILIAMS setup at VERA}

\begin{figure}[b!]
  \centering
    \includegraphics[width=0.8\textwidth]{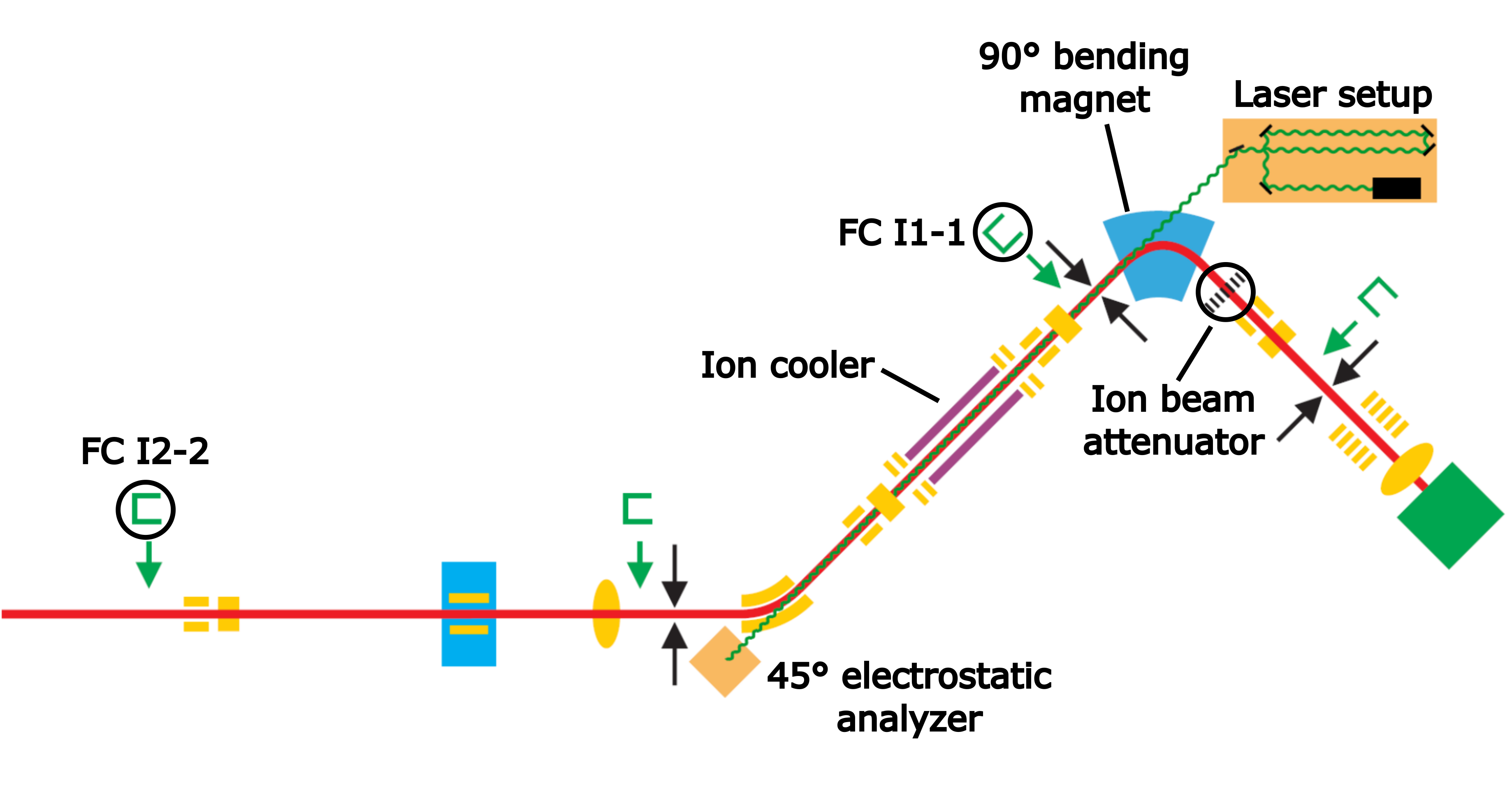}
    \caption{Schematics of the ILIAMS setup.}\label{fig:1}
\end{figure}

All the measurements presented in this work were performed with the VERA AMS facility at the University of Vienna \cite{Steier2004,Golser2017}, utilizing the injection line with the ILIAMS setup, whose schematic is presented in \autoref{fig:1}. This setup is extensively described in various works \cite{Martschini2017,Martschini2019,Martschini2022}. ILIAMS employs laser photodetachment for the suppression of the stable isobar of the radionuclide of interest. This technique uses a laser beam to detach the electron from the anion of the potentially interfering isobar without affecting the anion from the radionuclide of interest. Therefore, it is necessary to find a suitable elemental or molecular anion species that has a higher detachment energy (DE) for the radionuclide of interest than for the stable isobar, and to use photons with an energy between these two DEs.

In the ILIAMS injection line, a beam of negative ions with an energy of 30\,keV is produced from a Multi Cathode Source of Negative Ions by Cesium Sputtering (MC-SNICS). The mass of the ions to be injected into the ILIAMS ion cooler is selected by a 90° bending magnet. An attenuator, consisting on a perforated metal sheet, can be inserted between the ion source and the magnet to reduce the ion current injected into the cooler by a factor of up to 60 \cite{Lachner2019}. The ion cooler chamber is set to a negative potential so that the ions are decelerated to an energy lower than 100\,eV. Inside the cooler, ions collide with He gas, which further reduces their energy, and ions are transversely trapped by the potential from radio-frequency quadrupole electrodes, similar to a linear Paul trap. This cooling of the ions is required to confine the ions in the center of the cooler, and to extend their interaction with a collinearly overlapped laser beam. The cooling (slowing down) of the ions also increases the residence time of the ions inside the cooler and, therefore, enhances the suppression of the interfering isobar. Ions are longitudinally accelerated to energies slightly below 1\,eV to the exit of the cooler by the so-called "guiding electrodes" to prevent them from getting trapped in the cooler. At the exit, the ions are reaccelerated to an energy of 30\,keV and transported towards the accelerator of the VERA facility. Ion currents injected into the cooler were measured with a Faraday cup located after the magnet (FC I1-1 in \autoref{fig:1}). The extracted ion current was measured in a Faraday cup located after the electrostatic analyzer that directs the ion beam from the cooler to the point where ILIAMS beamline connects with the rest of the VERA facility (FC I2-2 in \autoref{fig:1}).

\begin{figure}[t!]
  \centering
    \includegraphics[width=0.8\textwidth]{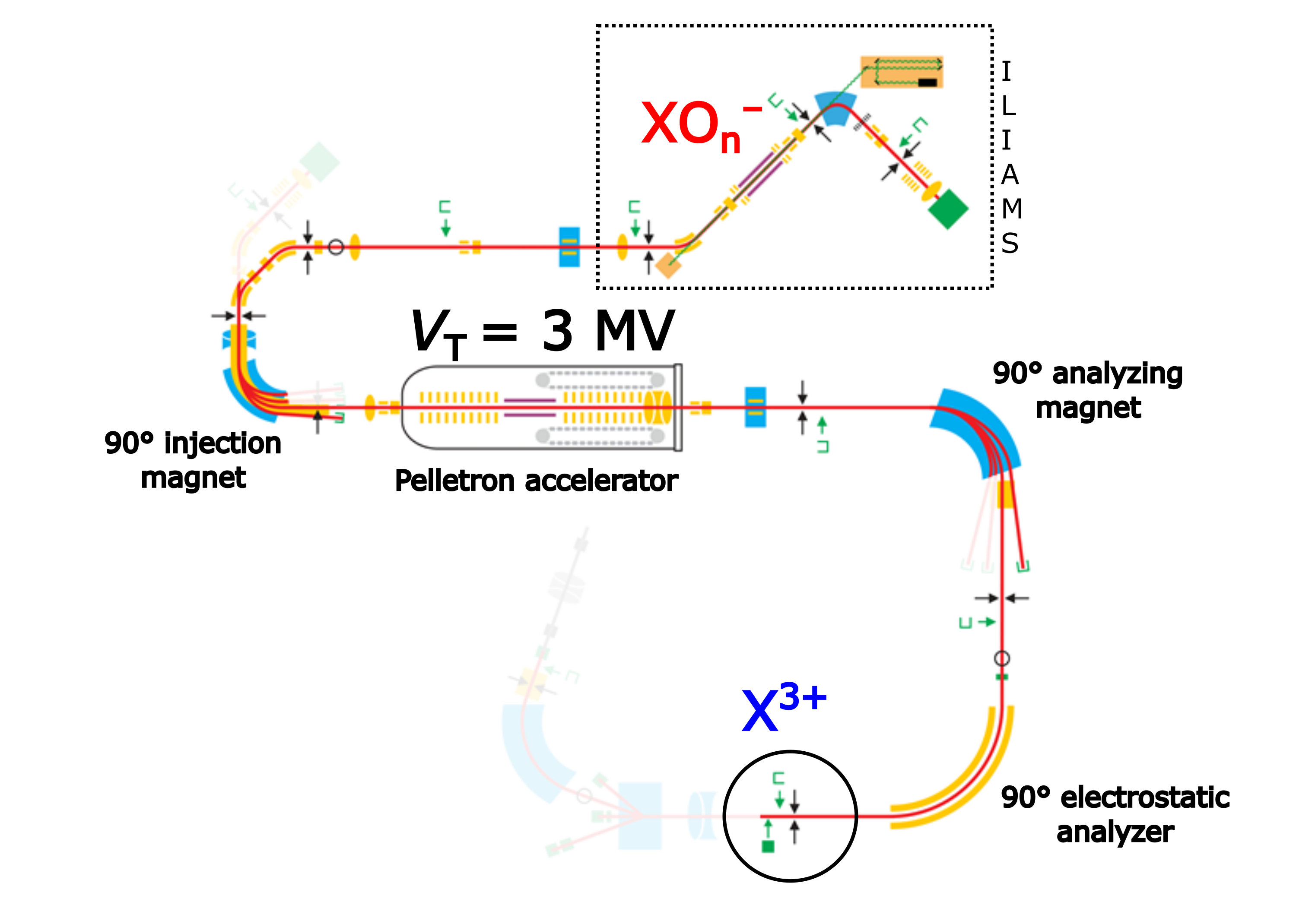}
    \caption{Basic schematic from the AMS setup for the measurement of \textsuperscript{91,94}Nb and \textsuperscript{93}Mo. In the case of X\,=\,\textsuperscript{91,94}Nb, the molecular ion injected into the cooler and into the accelerator is NbO\textsubscript{3}\textsuperscript{--}. In the case of X\,=\,\textsuperscript{93}Mo, the selected molecular ion is MoO\textsubscript{2}\textsuperscript{--}.}\label{fig:2}
\end{figure}

From this point, the procedure resembles a conventional AMS measurement at VERA. The mass of the negative ions to be injected into the accelerator is selected by a 90° bending magnet (injection magnet); the ions are accelerated in the pelletron accelerator, with terminal voltages up to 3\,MV, where the stripping process destroys the molecular ions. The mass/charge ratio on the high-energy side is selected using a 90° bending magnet (analyzer magnet) and a 90° electrostatic analyzer (ESA). The ions from the radionuclide of interest are detected by a gas ionization chamber (GIC) and the current of the ions from, at least, one of its stable isotopes is measured in a Faraday cup (FC). A basic schematic of the setup used in the \textsuperscript{91,94}Nb and \textsuperscript{93}Mo measurements at VERA, is shown in \autoref{fig:2}.

\subsection{Materials and lasers for the study of the suppression of stable isobars of \textsuperscript{91}Nb, \textsuperscript{94}Nb and \textsuperscript{93}Mo at ILIAMS}

Three different lasers were used for the studies presented here, with wavelengths of: 355\,nm (3.49\,eV photon energy, 100\,kHz repetition rate, AVIA LX, Coherent Inc.), 532\,nm (2.33\,eV, cw, VERDI V18, Coherent Inc.) and 637\,nm (1.95\,eV, Civillasers 15\,W Semiconductor Laser High Power Laser System, NaKu Technology Co., Ltd.). 

Materials used in this study were solely commercially available oxide materials of Nb, Mo, and Zr in powder form:
\begin{itemize}
    \item Nb\textsubscript{2}O\textsubscript{5} (Alfa Aesar Puratronic\textsuperscript{TM})
    \item MoO\textsubscript{3} (Alfa Aesar Puratronic\textsuperscript{TM})
    \item ZrO\textsubscript{2} (Alfa Aesar Puratronic\textsuperscript{TM})
\end{itemize}
In all these cases, the oxide powders were pressed into Al targets, without mixing the oxide material with any binder. The targets were placed into an Al sample wheel.

\section{Results and Discussion}\label{sec:results}

\subsection{Identification of suitable molecular systems}\label{subsec:identification}

Since the detachment energies of many molecular anions of the elements of interest here were not precisely known from literature, a preliminary survey to identify suitable molecular systems had to be conducted. The objective of this experiment was to determine: the optimal oxide anion species for \textsuperscript{91,94}Nb measurements at VERA, and the one for \textsuperscript{93}Mo measurements; as well as the required wavelength to suppress the analogous anions of \textsuperscript{91,94}Zr and \textsuperscript{94}Mo for \textsuperscript{91,94}Nb, and the analogous anion of \textsuperscript{93}Nb for \textsuperscript{93}Mo. For this, the ion currents of the oxide anions of interest extracted from the ion cooler were compared with and without the laser beam interacting inside the cooler for three different wavelengths. A summary of the results is shown in \autoref{tab:1}. Please note, however, that no mass selection is performed between the extraction from the cooler and the measurement of their current in FC I2-2, hence suppression values are only indicative and were more accurately determined for selected promising systems in the experiments in section \ref{subsec:nb}.

\begin{table}[t!]
    \caption{Ion current from the different oxide anions of Zr, Nb and Mo before and after the ion cooler depending on the laser used. The He pressure in the cooler was 3.9 Pa.}
    \begin{tabular}{lllllll}
    \hline
    \hline
        &  &  &  &  & \textbf{Ion current} & \textbf{Ion current}\\
        &  & \textbf{Target} & \textbf{Wavelength} & \textbf{Ion current} & \textbf{at FC I2-2} & \textbf{at FC I2-2}\\
        \textbf{Anion}  & \textbf{X\,=} & \textbf{material} & \textbf{(nm)} & \textbf{at FC I1-1 (nA)} & \textbf{laser off (nA)} & \textbf{laser on (nA)}\\
    \hline
    \hline
        \multirow{3}{*}{XO\textsuperscript{--}} & \textsuperscript{92}Zr & ZrO\textsubscript{2} & 637 & 6.8* & 2.57* & 0.01*\\\cline{2-7}
         & \textsuperscript{93}Nb & Nb\textsubscript{2}O\textsubscript{5} & 637 & 16* & 3.5* & 0.03*\\\cline{2-7}
         & \textsuperscript{92}Mo & MoO\textsubscript{3} & 637 & 3* & 1.24* & 0.14*\\
    \hline
    \hline
        \multirow{4}{*}{XO\textsubscript{2}\textsuperscript{--}} & \multirow{2}{*}{\textsuperscript{93}Nb} & \multirow{2}{*}{Nb\textsubscript{2}O\textsubscript{5}} & 637 & \multirow{2}{*}{16*} & \multirow{2}{*}{5.3*} & 0.2*\\
         &  &  & 532 &  &  & 0.01*\\\cline{2-7}
         & \multirow{2}{*}{\textsuperscript{92}Mo} & \multirow{2}{*}{MoO\textsubscript{3}} & 637 & \multirow{2}{*}{117} & \multirow{2}{*}{52} & 43\\
         &  &  & 532 &  &  & 0.3\\
    \hline
    \hline
         \multirow{6}{*}{XO\textsubscript{3}\textsuperscript{--}} & \multirow{2}{*}{\textsuperscript{92}Zr} & \multirow{2}{*}{ZrO\textsubscript{2}} & 532 & \multirow{2}{*}{2.3} & \multirow{2}{*}{1.0} & 0.63\\
         &  &  & 355 &  &  & 0.3\\\cline{2-7}
         & \multirow{2}{*}{\textsuperscript{93}Nb} & \multirow{2}{*}{Nb\textsubscript{2}O\textsubscript{5}} & 532 & \multirow{2}{*}{8.6*} & \multirow{2}{*}{3.9*} & 2.7*\\
         &  &  & 355 &  &  & 3.4*\\\cline{2-7}
         & \multirow{2}{*}{\textsuperscript{92}Mo} & \multirow{2}{*}{MoO\textsubscript{3}} & 532 & \multirow{2}{*}{8.4*} & \multirow{2}{*}{3.6*} & 4.0*\\
         &  &  & 355 &  &  & 0.01*\\
    \hline
    \hline
    \end{tabular}
    {*) During these measurements, the ion current was attenuated by the perforated metal sheet between the ion source and the magnet.}
    \label{tab:1}
\end{table}

Typical transmissions through the cooler for all these ion species are between 30\% and 50\%. (despite these losses, the overall detection efficiency is higher than at any high-energy AMS facility since abundant charge states like the 3+ can be used after the accelerator and the virtually isobar-free ion beam allows almost 100\% efficiency of the ROI in the detector). For the three elements, the monoxide anions (XO\textsuperscript{--}) provide high output currents, reaching more than 1\,\textmu A of \textsuperscript{93}NbO\textsuperscript{–-} in some cases. Nevertheless, all these anions suffered photodetachment even by their interaction with our laser with the lowest photon energy: the red laser, with a wavelength of 637\,nm. 

In the case of dioxide anions (XO\textsubscript{2}\textsuperscript{--}), the same 637\,nm laser effectively suppressed the current of NbO\textsubscript{2}\textsuperscript{--} by, at least, 1 order of magnitude, without severely affecting the one from MoO\textsubscript{2}\textsuperscript{--}. The \textsuperscript{92}MoO\textsubscript{2}\textsuperscript{--} output current from our ion source for targets prepared from MoO\textsubscript{3} is typically above 75\,nA. This related to a total \textsuperscript{nat}MoO\textsubscript{2}\textsuperscript{--} current above 500\,nA. Therefore we chose this measurement setup for \textsuperscript{93}Mo, where the suppression of \textsuperscript{93}Nb is required.

Injecting trioxide anions (XO\textsubscript{3}\textsuperscript{--}) into the cooler while overlapping the ion beam with the UV laser (355\,nm) suppresses the ZrO\textsubscript{3}\textsuperscript{--} and MoO\textsubscript{3}\textsuperscript{--} currents by, at least, one order of magnitude without severely affecting NbO\textsubscript{3}\textsuperscript{--}. Even when the \textsuperscript{93}NbO\textsubscript{3}\textsuperscript{--} current was almost two times lower than the ones for \textsuperscript{93}NbO\textsuperscript{--} and \textsuperscript{93}NbO\textsubscript{2}\textsuperscript{--}, it was still above 250\,nA. According to these results, we chose this setup for \textsuperscript{91}Nb measurements, where the suppression of \textsuperscript{91}Zr is required, and for \textsuperscript{94}Nb measurements, where the suppression of both \textsuperscript{94}Zr and \textsuperscript{94}Mo is required. 

The slight reduction of the NbO\textsubscript{3}\textsuperscript{--} current by the 355\,nm laser as well as the one for MoO\textsubscript{2}\textsuperscript{--} by the 637\,nm laser are attributed to the photodetachment of excited states of these ions. These ion currents did not decrease when increasing the buffer gas pressure. This would be expected if the ions in their ground state suffered photodetachment due to the increased residence time of the ions within the cooler \cite{Martschini2017,Martschini2019,Martschini2022,Lachner2019}. We relate the fact that the reduction of the NbO\textsubscript{3}\textsuperscript{--}  current is higher for the 532\,nm laser to the higher photon flux of this laser and the better overlap with the ion beam.

\subsection{Suppression of ZrO\textsubscript{3}\textsuperscript{--} and MoO\textsubscript{3}\textsuperscript{--} for AMS measurements of \textsuperscript{91}Nb and \textsuperscript{94}Nb}\label{subsec:nb}

While \textsuperscript{91}Nb has only one stable isobar, \textsuperscript{91}Zr, measurements of \textsuperscript{94}Nb require the suppression of two stable isobars: \textsuperscript{94}Zr and \textsuperscript{94}Mo. Therefore, to be able to measure any of these two isotopes of niobium at VERA, the first requirement is to achieve a sufficient suppression of ZrO\textsubscript{3}\textsuperscript{--} with ILIAMS.

\begin{figure}[t!]
  \centering
    \includegraphics[width=0.6\textwidth]{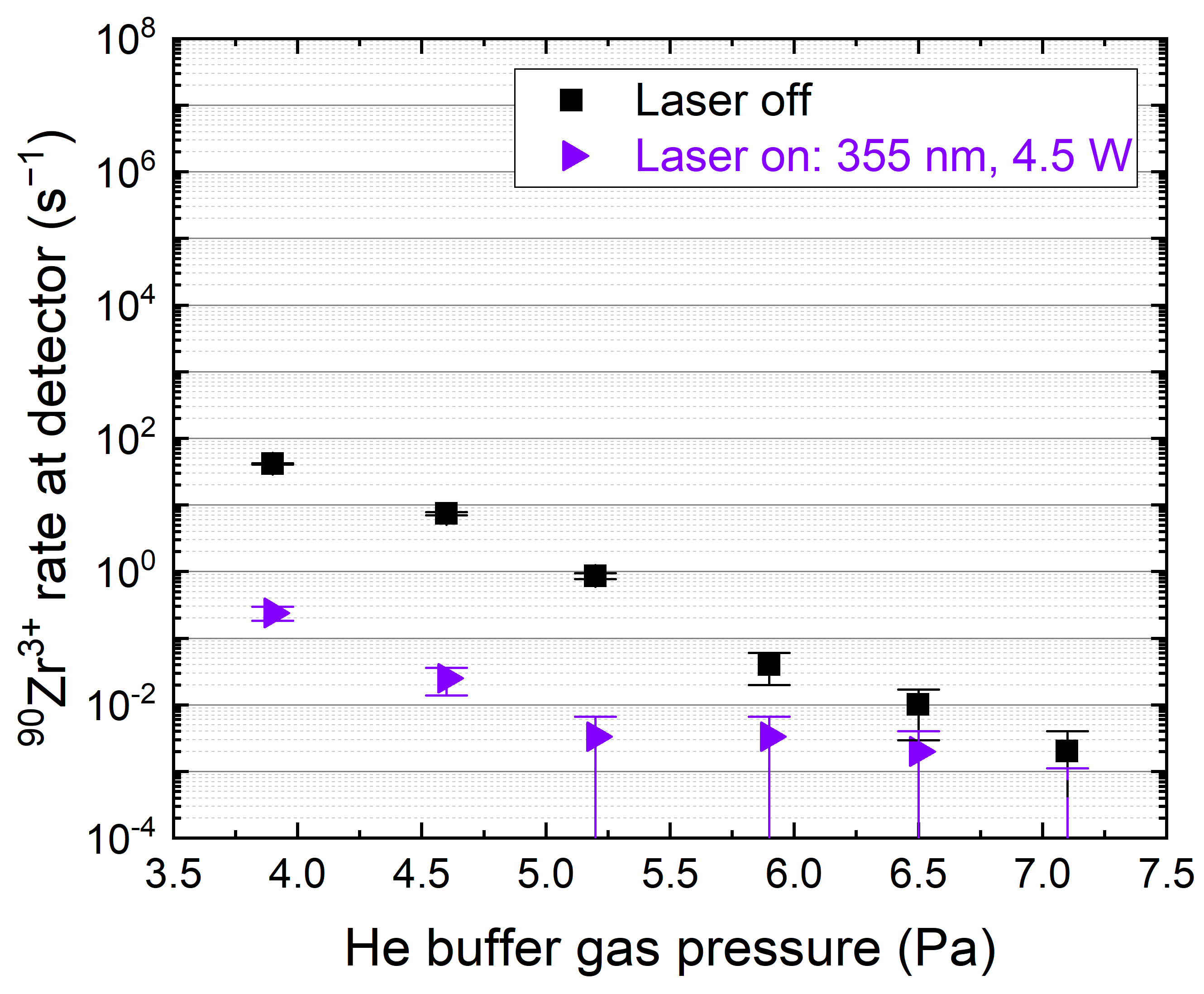}
    \caption{\textsuperscript{90}Zr\textsuperscript{3+} rate in the detector from a Nb\textsubscript{2}O\textsubscript{5} target as a function of the He buffer gas pressure with and without laser. \textsuperscript{90}ZrO\textsubscript{3}\textsuperscript{--} is injected into the cooler. Collisions with the gas already suppress ZrO\textsubscript{3}\textsuperscript{--}. The suppression is enhanced by the 355\,nm laser, leading to no \textsuperscript{90}Zr\textsuperscript{3+} counts observed in 900\,s with a He buffer pressure of 7.1\,Pa and the laser on.}\label{fig:zr-sup}
\end{figure}

As shown in \autoref{fig:zr-sup}, even without the laser, the \textsuperscript{90}Zr\textsuperscript{3+} rate in the detector decreased exponentially when increasing the He buffer gas pressure inside the cooler. This means that ZrO\textsubscript{3}\textsuperscript{--} is suppressed by collision with the buffer gas. An accurate suppression factor, therefore, could not be determined, since we can assume that ZrO\textsubscript{3}\textsuperscript{--} already experienced some suppression even for a He pressure in the cooler of 3.9\,Pa. The suppression is enhanced when the 355\,nm laser is overlapped with the ions in the cooler. Uncertainties here and in the following are always based on the square root of residual counts from the isobar induced background divided (statistical uncertainty) by the measurement time (for 0~counts, 1~count is assumed as uncertainty). With a He buffer gas pressure of 7.1\,Pa and the laser on, no \textsuperscript{90}Zr\textsuperscript{3+} counts were registered during a measurement time of 900\,s, corresponding to a lower limit for the suppression factor of 37000.

The \textsuperscript{93}Nb\textsuperscript{3+} current in the insertable Faraday cup before the detector was approximately 10\,nA during the experiment. The upper limit of 1~count in a measurement time of 900\,s corresponds to a \textsuperscript{90}Zr/\textsuperscript{93}Nb ratio of 5.3\,$\times$\,10\textsuperscript{--14}. Considering the isotopic abundances of the Zr isotopes, this would correspond to upper limits for the \textsuperscript{91}Zr/\textsuperscript{93}Nb and \textsuperscript{94}Zr/\textsuperscript{93}Nb isobar induced backgrounds of 1.2\,$\times$\,10\textsuperscript{--14} and 1.8\,$\times$\,10\textsuperscript{--14}, respectively.

\begin{figure}[t!]
  \centering
    \includegraphics[width=0.6\textwidth]{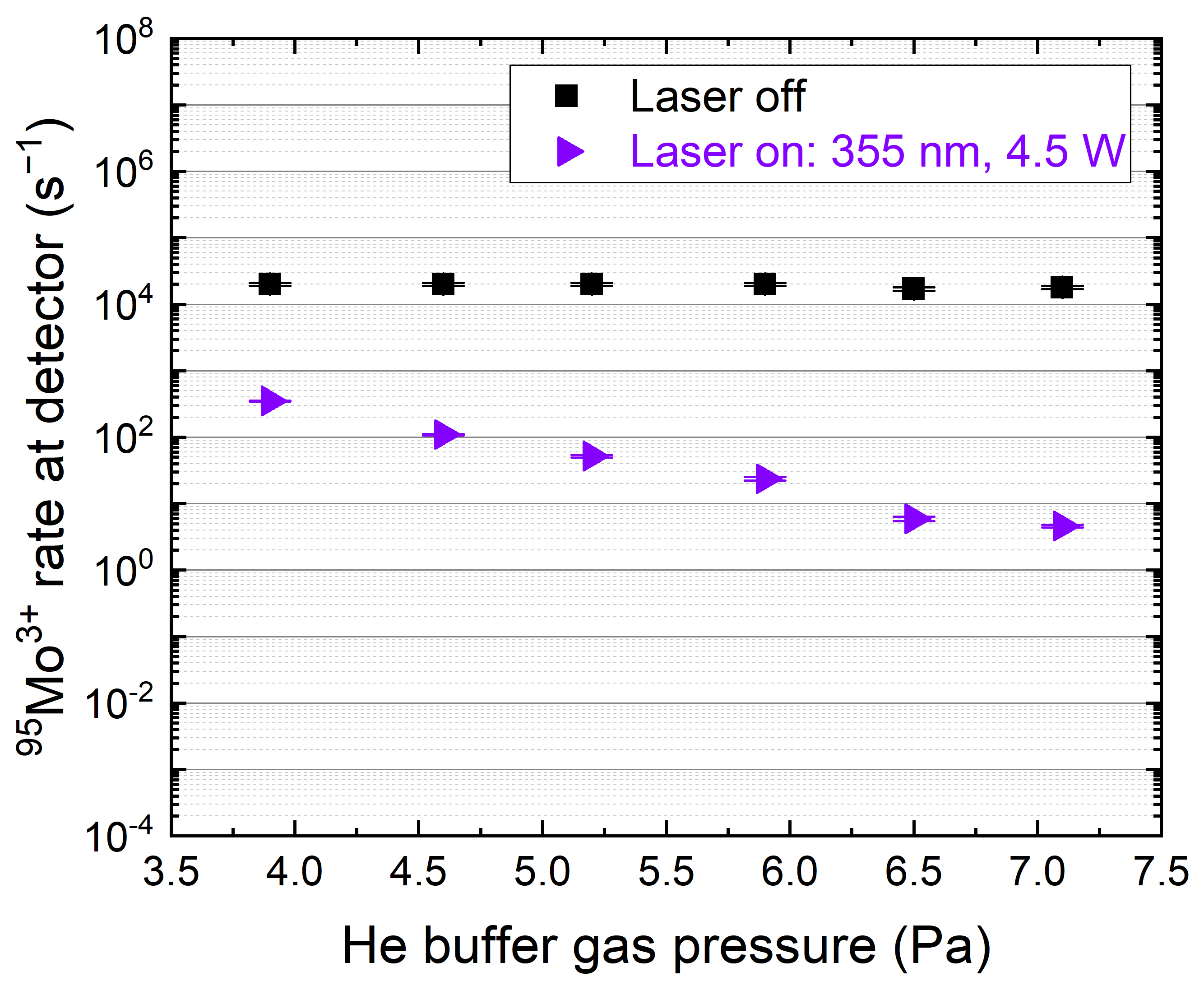}
    \caption{\textsuperscript{95}Mo\textsuperscript{3+} rate in the detector from a Nb\textsubscript{2}O\textsubscript{5} target as a function of the He buffer gas pressure with and without laser. \textsuperscript{95}MoO\textsubscript{3}\textsuperscript{--} is injected into the cooler, which is suppressed by a factor 4360 by the 355\,nm laser.}\label{fig:mo-sup}
\end{figure}

As shown in \autoref{fig:mo-sup}, the \textsuperscript{95}Mo\textsuperscript{3+} count rate, of 20000\,s\textsuperscript{--1}, remained unchanged when varying the He buffer gas pressure in the ion cooler. Hence, in contrast to ZrO\textsubscript{3}\textsuperscript{--}, MoO\textsubscript{3}\textsuperscript{--} is not suppressed by collisions with the buffer gas. However, a suppression by the 355\,nm laser increasing with buffer gas pressure was observed. This trend is expected for laser photodetachment at ILIAMS, since higher buffer gas pressures lead to longer residence times of the ions within the cooler, increasing the probability of interaction with the laser \cite{Martschini2017,Martschini2019,Martschini2022,Lachner2019}. At a He gas pressure in the cooler of 7.1\,Pa, laser photodetachment reduced the \textsuperscript{95}Mo\textsuperscript{3+} count rate to (4.59\,$\pm$\,0.21)\,s\textsuperscript{--1} (459~counts in 100\,s), corresponding to a suppression factor of 4360. This count rate translates to a \textsuperscript{95}Mo/\textsuperscript{93}Nb ratio of (2.21\,$\pm$\,0.10)\,$\times$\,10\textsuperscript{--10}, equivalent to a \textsuperscript{94}Mo/\textsuperscript{93}Nb isobar induced background of (1.28\,$\pm$\,0.06)\,$\times$\,10\textsuperscript{--10}.

For the foils irradiated during the DTE2 campaign at the JET reactor with a molybdenum content above 1\%, assuming that 50\,mg of sample material and 1\,mg of Nb carrier are used, the \textsuperscript{91}Nb/\textsuperscript{93}Nb and \textsuperscript{94}Nb/\textsuperscript{93}Nb ratios to be measured at VERA would be in the order of 10\textsuperscript{-–10} and 10\textsuperscript{-–11}, respectively. This means that \textsuperscript{91}Zr and \textsuperscript{94}Zr interferences should not limit the \textsuperscript{91,94}Nb measurement at VERA, provided that chemical sample preparation can reduce the zirconium content to levels similar to the sputter targets used in this study. On the other hand, to allow the measurement of their \textsuperscript{94}Nb concentrations, the suppression of \textsuperscript{94}Mo needs to be improved. This could involve studying the dependence of this suppression on various ion cooler parameters. An example would be the study of the dependence of the suppression factor on the energy with which the ions are injected into the ion cooler.

\subsection{Suppression of NbO\textsubscript{2}\textsuperscript{--} for AMS measurements of \textsuperscript{93}Mo}\label{subsec:mo}

The dioxide system was closely studied on MoO\textsubscript{3} sputter targets by comparing the ion current of \textsuperscript{92}Mo\textsuperscript{3+} in front of the detector when injecting \textsuperscript{92}MoO\textsubscript{2}\textsuperscript{-–} with the \textsuperscript{93}Nb\textsuperscript{3+} rate in the detector when injecting \textsuperscript{93}NbO\textsubscript{2}\textsuperscript{-–}, respectively.

\begin{figure}[t!]
  \centering
    \includegraphics[width=0.6\textwidth]{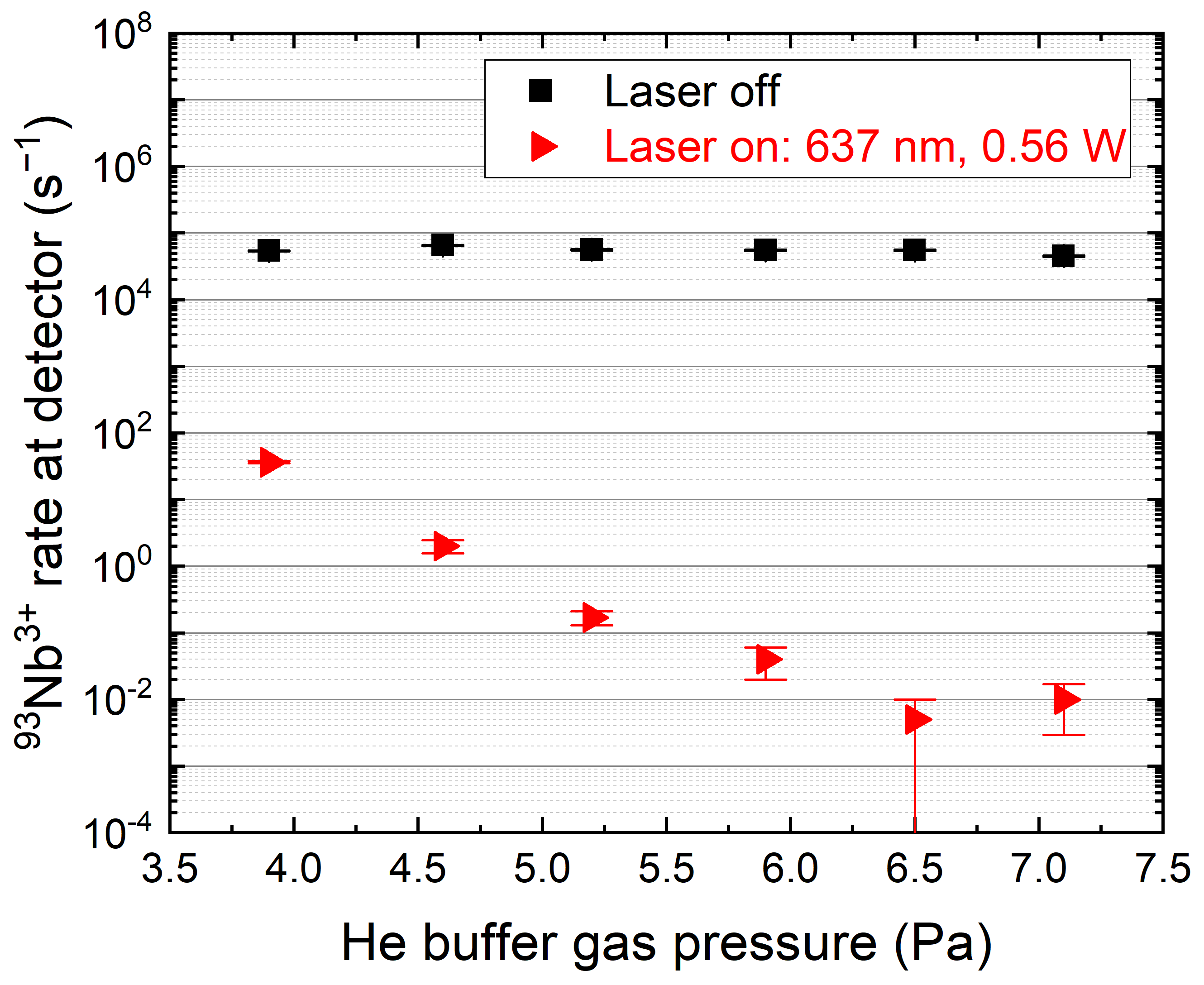}
    \caption{\textsuperscript{93}Nb\textsuperscript{3+} rate in the detector from a MoO\textsubscript{3} target as a function of the He buffer gas pressure with and without laser. The \textsuperscript{93}NbO\textsubscript{2}\textsuperscript{--} anion is injected into the cooler, which is suppressed 6 orders of magnitude by the 637\,nm laser.}\label{fig:nb-sup}
\end{figure}

\begin{figure}[h!]
  \centering
    \includegraphics[width=0.8\textwidth]{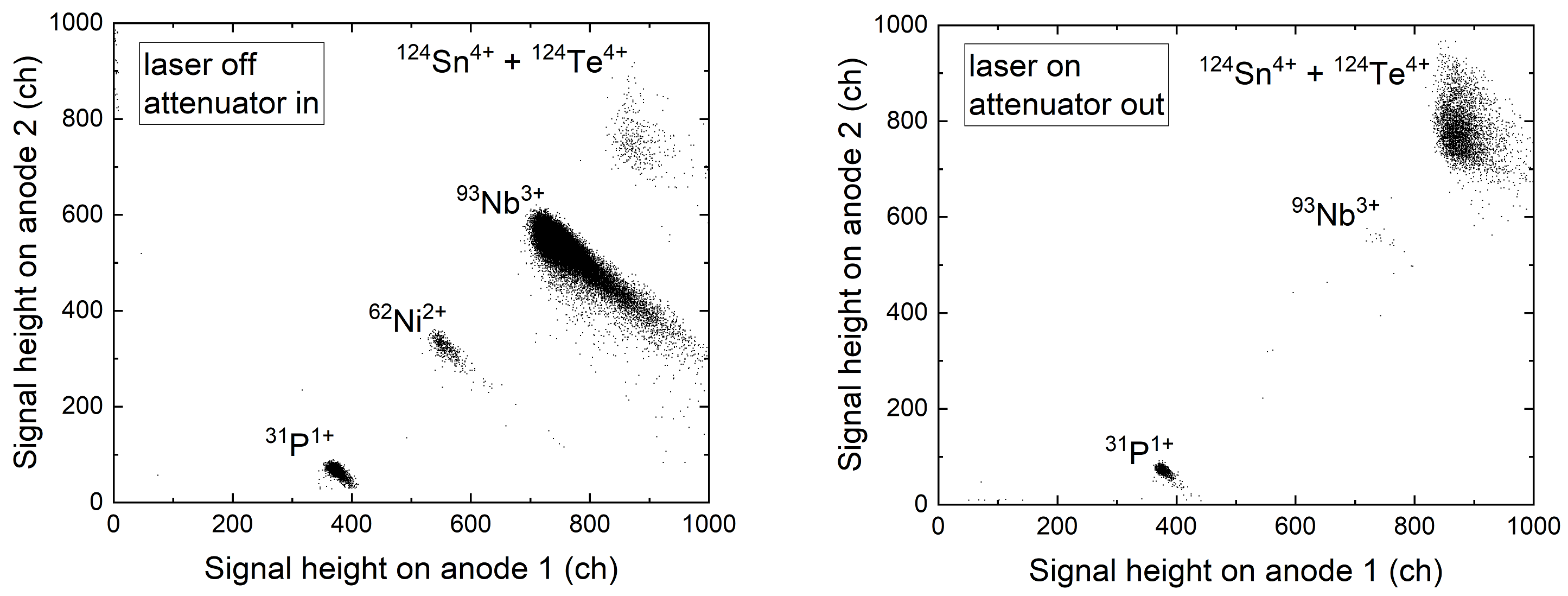}
    \caption{Typical spectra in the GIC detector for the \textsuperscript{93}MoO\textsubscript{2}\textsuperscript{–-}\,$\rightarrow$\,\textsuperscript{93}Mo\textsuperscript{3+} setup at VERA on a target prepared from \textsuperscript{93}Mo-free MoO\textsubscript{3} depending on if the 637\,nm laser is off (left) or on (right). In the first case, the ion beam attenuator had to be inserted to avoid the saturation of the detector. The measurement time was 960\,s in the first case and 600\,s in the second.}\label{fig:6}
\end{figure}

As illustrated in \autoref{fig:nb-sup}, NbO\textsubscript{2}\textsuperscript{--} was not suppressed by collisions with the He buffer gas. However, the electron photodetachment by the 637\,nm laser significantly suppressed the \textsuperscript{93}Nb\textsuperscript{3+} count rate, from 55000\,s\textsuperscript{--1} down to an upper limit of the rate of 0.02\,s\textsuperscript{--1}, corresponding to a suppression factor $>$2.7\,$\times$\,10\textsuperscript{6}. During the experiment, the \textsuperscript{92}Mo\textsuperscript{3+} ion current measured in the Faraday cup before the detector was 5.6\,nA, resulting in an upper limit for the \textsuperscript{93}Nb/\textsuperscript{nat}Mo isobar induced background of 2.5\,$\times$\,10\textsuperscript{--13}. Examples of the typical spectra in the GIC detector, depending on if the 637\,nm laser is off or on are presented in \autoref{fig:6}.

For the foils irradiated during the DTE2 campaign at the JET reactor with a molybdenum content above 1\%, our preliminary calculations indicate that this interference should be three orders of magnitude below the expected \textsuperscript{93}Mo/\textsuperscript{nat}Mo. Furthermore, the strong suppression of the \textsuperscript{93}Nb interference potentially allow the measurement of \textsuperscript{93}Mo even in samples from reduced activation steels, where molybdenum is present only in trace amounts.

\section{Conclusions and prospects}\label{sec:conclu}

ZrO\textsubscript{3}\textsuperscript{--} is strongly suppressed by collisions with the He buffer gas in the ion cooler of the ILIAMS setup alone. This opens the possibility of performing \textsuperscript{91}Nb measurements at VERA without usage of laser photodetachment. If required, this suppression can be enhanced by overlapping the 355\,nm laser with the ion beam within the cooler. This laser also suppresses MoO\textsubscript{3}\textsuperscript{--}, as required for \textsuperscript{94}Nb measurements, by a factor 4360.

NbO\textsubscript{2}\textsuperscript{--} is suppressed by a factor $>$2.7\,$\times$\,10\textsuperscript{6} with the 637\,nm laser, potentially enabling measurements of the \textsuperscript{93}Mo concentration in samples where molybdenum is present only in trace amounts.

Work in the near future will focus on optimizing the suppression of interfering isobars, particularly the \textsuperscript{94}Mo interference in \textsuperscript{94}Nb measurements, and testing the developed setups with chemically processed samples and the actual radionuclides \textsuperscript{91}Nb, \textsuperscript{94}Nb and \textsuperscript{93}Mo. Efforts will also be directed towards the production and measurement of reference materials for these three radionuclides.

\section*{Acknowledgements}

The authors acknowledge financial support of the ILIAMS research activities by “Investitionsprojekte” of the University of Vienna. The authors are indebted to the rest of students and staff of the Isotopes Physics group of the University of Vienna for their continuous support.

\section*{Declaration of generative AI and AI-assisted technologies in the writing process}

During the preparation of this work the author(s) used Acrobat AI Assistant in order to correct the language an readability of the text written by the author(s). After using this tool/service, the author(s) reviewed and edited the content as needed and take(s) full responsibility for the content of the publication.

\bibliographystyle{elsarticle-num}
\bibliography{biblio}

%%% End document %%%
\end{document}